# Generalized Hooke's law for isotropic second gradient materials

By F. dell'Isola[1,3], G. Sciarra[2] and S. Vidoli[1,*]

[1]*Dipartimento di Ingegneria Strutturale e Geotecnica, and* [2]*Dipartimento di Ingegneria Chimica Materiali Ambiente, Università di Roma La Sapienza, via Eudossiana 18, 00184 Rome, Italy*
[3]*Laboratorio di Strutture e Materiali Intelligenti, Università di Roma La Sapienza, Ala Nord, Palazzo Caetani, 04012 Cisterna di Latina, Italy*

In the spirit of Germain the most general objective stored elastic energy for a second gradient material is deduced using a literature result of Fortuné & Vallée. Linear isotropic constitutive relations for stress and hyperstress in terms of strain and strain-gradient are then obtained proving that these materials are characterized by seven elastic moduli and generalizing previous studies by Toupin, Mindlin and Sokolowski. Using a suitable decomposition of the strain-gradient, it is found a necessary and sufficient condition, to be verified by the elastic moduli, assuring positive definiteness of the stored elastic energy. The problem of warping in linear torsion of a prismatic second gradient cylinder is formulated, thus obtaining a possible measurement procedure for one of the second gradient elastic moduli.

## 1. Introduction

Three-dimensional Cauchy continua represent a mathematical model suitable to describe many phenomena occurring in bodies which are deformed because of external actions. However, some deformation problems in linear elasticity have solutions which present singularities; the Flamant–Boussinesq problem and the Kelvin problem are two well-known examples (Georgiadis & Anagnostou 2008). It is therefore clear that when highly concentrated stress occur, a more detailed description of deformation phenomena is required. On the other hand, Cosserat & Cosserat (1909) already generalized Cauchy model to describe continuous media in which microrotations play a relevant role. Most recently, Cosserat models were used by many authors (e.g. Ehlers & Volk 1998) to describe granular materials, fluid saturated porous media or soils. Stress concentration phenomena revealed to be of importance in plasticity and fracture mechanics: in proximity of the region where plastic deformations take place Cauchy models are not able to catch some relevant phenomena (e.g. Deborst & Sluys 1991). Shear bands, transition regions between elastic and plastic regimes, crack initiations are some among them which require more encompassing models (e.g. Unger & Aifantis 2000). Similarly, when modelling the constitutive behaviour of two-phase composites, Drugan & Willis (1996) found that the leading-order correction to a homogeneous constitutive equation involves a term proportional to the second gradient of the ensemble average of strain.

*Author for correspondence (stefano.vidoli@uniroma1.it).

A possible generalization of Cauchy models including those proposed by Cosserat & Cosserat (1909) can be found in Toupin (1962), Mindlin (1964) and Germain (1973). In these papers, continuum mechanics is framed in a setting where stored (deformation) energy depends not only on strain, but also on strain gradient. The more general continua, thus introduced, are called second gradient continua by Germain or first strain-gradient continua by Mindlin. In these approaches, the concept of contact actions needs more general treatment than the one due to Cauchy. Indeed, second gradient continua show surface contact actions of two different types: contact surface forces and contact surface double-forces. Moreover, when Cauchy cuts present edges, then also contact line forces arise (dell'Isola & Seppecher 1997). To analyse the dependence of stored energy on strain gradient and the relationship between this energy and contact actions, it is necessary to establish some properties of third-order tensors.

Each third-order tensor $S_{ijk}$, symmetric with respect to its first two indices, can be uniquely decomposed as the sum of a third-order tensor, completely symmetric with respect to all permutations of its indices, and a 'sym-skew' third-order tensor, (skew-symmetric) symmetric with respect to its (last) first two indices. We refer to equations (2.10) and (2.11) for a precise definition.

The strain gradient is a third-order tensor which is symmetric with respect to its first two indices: therefore, it can be decomposed into its completely symmetric and sym-skew parts. In Toupin (1962) and Mindlin (1964), the stored energy depends only on the sym-skew part of the strain gradient, while the complete dependence is assumed in Germain (1973). As a consequence, the contact double forces reduce in Toupin (1962) and Mindlin (1964) to couple stresses, while no such limitation is assumed in Germain (1973), where Piola–Kirchhoff stress and hyperstress tensors are introduced but no constitutive assumption for the stored energy is discussed.

In the present paper we deduce, using a variational principle, equilibrium equations for second gradient materials in terms of the stored energy density, thus obtaining the corresponding representation of Piola–Kirchhoff stress and hyperstress. No small deformation assumption is necessary to this aim and, therefore, the obtained equations are valid in finite elasticity. Using the representation theorems of Suiker & Chang (2000), the most general homogeneous quadratic isotropic constitutive relation for second gradient stored (deformation) energy is found. As a consequence, constitutive relations for Piola–Kirchhoff stress and hyperstress in terms of strain and strain-gradient are obtained: these are generalized Hooke's laws valid for isotropic second gradient materials. For these materials, together with standard Lamé moduli, five more moduli are needed.

Subsequently, it is deduced a necessary and sufficient condition assuring that the stored energy is positive definite: to this aim it has been necessary to introduce a further decomposition of the strain gradient. The constitutive equations given by Mindlin (1964) and used by Sokolowski (1970) are found as a particular case when three out of the five aforementioned moduli vanish.

One of the most relevant problem to be addressed, when new constitutive equations are introduced, concerns the design of experimental procedures to apply for getting measurements of newly introduced moduli. To this aim, in the last section, the closed-form solution for the linear torsion of a prismatic, hollow, circular, second gradient cylinder is found which allows for the determination of the torsional rigidity in term of the standard shear modulus and of one specific second gradient modulus. For a generic cross section the problem of warping in linear torsion is formulated; the differences with a study due to Sokolowski (1970) are discussed.

## 2. Equilibrium equations for second gradient materials

Second gradient theories of deformable bodies are characterized by requiring the volume density of the virtual internal work $W^{\text{int}}$ to depend linearly on virtual displacement field $\delta u$ and on its first and second spatial gradients, see Toupin (1962), Mindlin (1964) and Germain (1973)

$$W^{\text{int}}(\delta u) = -\int_{\mathcal{D}} \left( \varsigma_\alpha \delta u_\alpha + \Sigma_{\alpha\beta} \delta u_{\alpha,\beta} + \Pi_{\alpha\beta\gamma} \delta u_{\alpha,\beta\gamma} \right); \tag{2.1}$$

here and in what follows we use commas to denote partial derivatives.

The principle of virtual work is therefore assumed to have the following form:

$$W^{\text{int}}(\delta u) + W^{\text{ext}}(\delta u) = 0, \tag{2.2}$$

where the equality is assumed to be valid for every sufficiently smooth virtual displacement field $\delta u$; and $\mathcal{D}$ is the Eulerian domain occupied by the considered body. From now on Greek indices refer to coordinates with respect to a fixed orthonormal frame in the Euclidean space and Einstein convention on repeated indices is used. The representation assumed for $W^{\text{int}}$ is made plausible by the Schwartz (1963) representation theorem for continuous linear functionals. Simple successive applications of Gauss theorem imply that the external work $W^{\text{ext}}$ has the following form (Seppecher 1989):

$$W^{\text{ext}}(\delta u) = \int_{\mathcal{D}} b_\alpha \delta u_\alpha + \int_{\partial \mathcal{D}} (t_\alpha \delta u_\alpha + \tau_\alpha \delta u_{\alpha,\beta} m_\beta) + \sum_h \int_{\mathcal{E}_h} f_\alpha \delta u_\alpha, \tag{2.3}$$

where $\partial \mathcal{D}$ is the boundary of $\mathcal{D}$, assumed to be differentiable almost everywhere; and $\mathcal{E}_h$ is the $h$th edge of $\partial \mathcal{D}$, if any. On the curve $\mathcal{E}_h$ the normal $m$ to the boundary $\partial \mathcal{D}$ suffers a jump.

As a consequence, the introduced second gradient internal actions,[1] the stress $\Sigma$ and the hyperstress $\Pi$, balance not only bulk forces $b$ and surface tractions $t$, but also surface double-forces $\tau$ and tractions per unit line $f$ (Germain 1973).

In this paper, the existence of a regular Lagrangian stored (deformation) energy density $\psi$ is assumed. Moreover, we assume the variation of the stored energy to equal the opposite of the internal work $W^{\text{int}}$ once this last has been expressed in terms of Lagrangian quantities. The consequent Lagrangian representation of the principle of virtual work allows us to determine the searched equilibrium equations.

---

[1] Zero-th order stress $\zeta$ has to vanish because of the invariance under change of observer of the internal work, see for instance Seppecher (1989).

## (a) Kinematics

A Lagrangian description of motion is adopted for second gradient solids. Let us consider the reference configuration $\mathcal{D}_0$ of the body as a sufficiently smooth domain of the Euclidean ambient space $\mathcal{E}$. Placement of material particles are determined by the bijective function $\chi : \mathcal{D}_0 \to \mathcal{E}$ mapping any material particle in its current position. Moreover, $\mathcal{D} := \chi(\mathcal{D}_0)$ and the displacement field is defined, as usual, $u_\alpha(X) := \chi_\alpha(X) - T_{\alpha i} X_i$, for $X \in \mathcal{D}_0$. Here and in what follows, Latin and Greek indices denote respectively the Lagrangian and Eulerian components with respect to fixed orthonormal frames in the reference and current configurations; moreover, the invertible matrix $T$ represents the parallel transport from coordinates in $\mathcal{D}_0$ to coordinates in $\mathcal{D}$; for the sake of simplicity, here we refer to both $\mathcal{D}_0$ and $\mathcal{D}$ in the same orthonormal coordinate system. Thus, $T_{\alpha i} = \delta_{\alpha i}$ with $\delta$ the Kronecker delta; the distinction between Lagrangian (Latin) and Eulerian (Greek) components is, therefore, not essential. However, this distinction will be useful when discussing the objectivity of the stored deformation energy, which involves changes of reference in the current configuration.

To deduce equilibrium equations from a variational principle the varied displacement map $u^*$ needs to be introduced, for all $X \in \mathcal{D}_0$:

$$u^*_\alpha(X_i) := \chi_\alpha(X_i) - T_{\alpha i} X_i + \delta u_\alpha(X_i), \qquad (2.4)$$

where $\delta u$ represents an arbitrary variation of the function $u$. The physical meaning of this variation is well-known in mechanics, and stands as the aforementioned virtual displacement. When no misunderstanding is possible, we will denote with the same symbol $\delta u$ and the corresponding Eulerian field $\delta u \circ \chi^{-1}$. It is not the purpose of this paper to determine the regularity assumptions for $\chi$ or $u$ which guarantee the existence and uniqueness of the deformation problem. Therefore we will limit ourselves to assume that sufficiently smooth placement maps are considered.

We denote the deformation tensor $F_{\alpha i} := \chi_{\alpha,i}$; here, $i$ means the direction in the reference configuration along which the derivative is computed. Requiring that $J := \det F > 0$, the polar decomposition $F_{\alpha i} = R_{\alpha,\beta} \delta_{\beta j} U_{ji}$ holds true, with $R \in \mathrm{Orth}^+$, the proper rotation, and $U \in \mathrm{Sym}^+$, the pure stretch. Here, $\mathrm{Orth}^+$ represents the group of orthogonal matrices with positive determinant; similarly, $\mathrm{Sym}^+$ is the cone of positive definite symmetric matrices. The mixed second-order tensor $\bar{R}$ with components $\bar{R}_{\alpha j} := R_{\alpha\beta} \delta_{\beta j}$ is usually labelled as the rotation component of $F$.

Finally, the Green–Lagrange strain tensor

$$E_{ik} = (F_{\alpha i} F_{\alpha k} - \delta_{ik})/2 = (U_{ij} U_{jk} - \delta_{ik})/2, \qquad (2.5)$$

is introduced to measure deformations with respect to the reference configuration.

Since $F$ is the Lagrangian gradient of the placement map $\chi$ it should satisfy the following compatibility (local integrability) conditions

$$0 = (\mathrm{curl}\, F)_{i\alpha} = -\varepsilon_{ijk} F_{\alpha j,k} = -\varepsilon_{ijk} \bar{R}_{\alpha l,k} U_{lj} - \varepsilon_{ijk} \bar{R}_{\alpha l} U_{lj,k}. \qquad (2.6)$$

Multiplying by $\bar{R}_{\alpha f}$ one gets the following:

$$0 = \varepsilon_{ijk}\bar{R}_{\alpha f}\bar{R}_{\alpha l,k}U_{lj} + \varepsilon_{ijk}U_{fj,k}, \tag{2.7}$$

where the third-order tensor $\bar{R}_{\alpha f}R_{\alpha l,k}$ represents the Lagrangian pull-back of the gradient of rotation (Lagrangian gradient of rotation, for the sake of conciseness). It is skew-symmetric with respect to $f$ and $l$ indices, therefore, it has at most nine independent components. Indeed, following Fortuné & Vallée (2001), the Lagrangian gradient $F$ is locally integrable and, therefore, equation (2.7) is identically verified, if

$$\bar{R}_{\alpha f}\bar{R}_{\alpha l,k} = \varepsilon_{flm}\Lambda_{mk}, \tag{2.8}$$

holds with

$$\Lambda_{mk} := \frac{1}{\det U}\left(U_{ml}(\text{curl } U)_{nl} - \frac{1}{2}U_{ij}(\text{curl } U)_{ij}\delta_{mn}\right)U_{nk}. \tag{2.9}$$

Equations (2.8) and (2.9) guarantee the Lagrangian gradient of rotation, and consequently $\nabla F$, to be represented in terms of the stretch $U$ and its gradient $\nabla U$. Here, and in the following, $\nabla$ denotes the Lagrangian gradient, i.e. gradient with respect to material coordinates. This result will be useful in §2b, when specifying the restrictions on the stored energy coming from objectivity requirements.

In what follows we apply to the strain gradient $E_{ij,k}$ a suitable decomposition valid for every third-order tensor symmetric with respect to the first two indices. This reads as follows:

$$K_{ijk} = \tilde{K}_{ijk} + \frac{1}{3}(\varepsilon_{jkl}\hat{K}_{li} + \varepsilon_{ikl}\hat{K}_{lj}), \tag{2.10}$$

where $\tilde{K}$ is a completely symmetric third-order tensor and $\hat{K}$ is a deviatoric second-order tensor, defined by,

$$\tilde{K}_{ijk} := \frac{1}{3}(K_{ijk} + K_{jki} + K_{kij}), \qquad \hat{K}_{li} := \varepsilon_{ljk}K_{ijk}, \tag{2.11}$$

and $\epsilon_{ijk}$ is the Levi-Civita alternator. The decomposition (2.10) is direct, i.e. the completely symmetric part $\tilde{K}$ is orthogonal to the sym-skew part of $K$, i.e. $(\varepsilon_{jkl}\hat{K}_{li} + \varepsilon_{ikl}\hat{K}_{lj})/3$. In Germain (1973) a similar decomposition was used for the second gradient of the velocity field, a third-order tensor symmetric with respect to the last two indices.

### (b) Objective stored elastic energy

Hyperelastic materials are constitutively characterized, in the context of first gradient theories, by considering a stored energy function $\psi$ (per unit volume) depending on the deformation $F$. In the framework of second gradient theories, their constitutive characterization is obtained by considering a stored energy depending both on the deformation tensor $F$ and its gradient $\nabla F$.

In classical elasticity the stored energy function $\psi$ depends just on stretch $U$, in order to be an objective function of the deformation $F$. Indeed, objectivity of $\psi$ in that case means

$$\psi(F_{\alpha i}) = \psi(Q_{\beta\alpha}F_{\alpha i}), \tag{2.12}$$

for every orthogonal tensor $Q \in \text{Orth}$. Considering in particular $Q_{\beta\alpha} = R_{\alpha\beta}$ one obtains

$$\psi(F_{\alpha i}) = \psi(R_{\alpha\beta} R_{\alpha\gamma} \delta_{\gamma j} U_{ji}) = \psi(\delta_{\beta j} U_{ji}) = \hat{\psi}(U_{ji}).$$

Similarly, when the stored energy depends on both the deformation $F$ and its gradient $\nabla F$, the objectivity condition implies,

$$\psi(F_{\alpha i}, F_{\alpha i,j}) = \psi(Q_{\beta\alpha} F_{\alpha i}, Q_{\beta\alpha} F_{\alpha i,j}), \tag{2.13}$$

for every orthogonal tensor $Q$. When considering $Q_{\beta\alpha} = R_{\alpha\beta}$ and according to equation (2.8), the second argument of the right-hand side in equation (2.13) becomes

$$\delta_{\beta k} Q_{\beta\alpha} F_{\alpha i,j} = \delta_{\beta k} R_{\alpha\beta}[(R_{\alpha\gamma} \delta_{\gamma l})_{,j} U_{li} + R_{\alpha\gamma} \delta_{\gamma l} U_{li,j}] = \bar{R}_{\alpha k} \bar{R}_{\alpha l,j} U_{li} + U_{ki,j}$$

$$= \varepsilon_{klm} \Lambda_{mj} U_{li} + U_{ki,j}. \tag{2.14}$$

Since, by equation (2.9), $\Lambda$ is univocally determined in terms of the stretch $U$ and its gradient $\nabla U$, every stored-energy functional satisfying (2.13) can be written in the form

$$\psi(F_{\alpha i}, F_{\alpha i,j}) = \hat{\psi}(U_{ik}, U_{ik,j}). \tag{2.15}$$

An equivalent form of (2.15) is also given in terms of the Green–Lagrange strain $E$, which will be used in the following when developing the variational formulation:

$$\psi(F_{\alpha i}, F_{\alpha i,j}) = \tilde{\psi}(E_{ik}, E_{ik,j}). \tag{2.16}$$

Using the decomposition (2.10) for the third-order tensor $E_{ik,j}$ the following dependence of $\psi$ can equivalently be considered:

$$\psi(F_{\alpha i}, F_{\alpha i,j}) = \bar{\psi}(E_{ik}, \tilde{E}_{ikj}, \hat{E}_{ik}). \tag{2.17}$$

### (c) Global equilibrium equations in Eulerian form

The global equilibrium equations for the second gradient solid are obtained requiring the external work (2.3) to vanish on every kinematically admissible rigid displacement field

$$\delta u_\alpha(x) = w_\alpha + \Omega_{\alpha\beta} x_\beta.$$

Here, $w$ is a constant vector representing the virtual translation, $\Omega$ is a constant skew-symmetric second-order tensor representing the angular virtual displacement and $x$ denotes the current place. One easily obtains the following:

$$0 = w_\alpha \left( \int_\mathcal{D} b_\alpha + \int_{\partial\mathcal{D}} t_\alpha + \sum_h \int_{\mathcal{E}_h} f_\alpha \right)$$

$$+ \Omega_{\alpha\beta} \left( \int_\mathcal{D} b_\alpha x_\beta + \int_{\partial\mathcal{D}} (t_\alpha x_\beta + \tau_\alpha m_\beta) + \sum_h \int_{\mathcal{E}_h} f_\alpha x_\beta \right). \tag{2.18}$$

The arbitrariness of $w$ and $\Omega$ leads to the global equilibrium equations of momentum and moment of momentum

$$\int_{\mathcal{D}} b_\alpha + \int_{\partial \mathcal{D}} t_\alpha + \sum_h \int_{\mathcal{E}_h} f_\alpha = 0, \qquad (2.19)$$

$$\int_{\mathcal{D}} (b_\alpha x_\beta - b_\beta x_\alpha) + \int_{\partial \mathcal{D}} (t_\alpha x_\beta - t_\beta x_\alpha + \tau_\alpha m_\beta - \tau_\beta m_\alpha) + \sum_h \int_{\mathcal{E}_h} (f_\alpha x_\beta - f_\beta x_\alpha) = 0. \qquad (2.20)$$

Note that the balance of momentum (2.19) involves not only bulk forces and tractions, but also contact edge-forces; on the other hand, a non-trivial contribution to the balance of moment of momentum (2.20) comes from skew-symmetric surface couples $\tau_\alpha m_\beta - \tau_\beta m_\alpha$ concentrated on the boundary.

### (d) Local equilibrium equations in Lagrangian form

The local equilibrium equations for a second gradient solid are obtained from the principle of virtual work (2.2) by standard localization arguments. In particular, these equations will be given in terms of Piola–Kirchhoff stress and hyperstress fields defined as suitable partial derivatives of the stored energy function. We find it useful to represent in Lagrangian description the internal work $W^{\text{int}}$ introduced in (2.1)

$$-\tilde{W}^{\text{int}} = \int_{\mathcal{D}_0} (s_i \delta u_i + S_{ij} \delta E_{ij} + P_{ijk} \delta E_{ij,k}). \qquad (2.21)$$

Using the identities

$$\left. \begin{aligned} \delta E_{ij} &= \frac{1}{2}(\delta F_{\alpha i} F_{\alpha j} + F_{\alpha i} \delta F_{\alpha j}), \\ \delta E_{ij,k} &= \frac{1}{2}(\delta F_{\alpha i,k} F_{\alpha j} + \delta F_{\alpha j,k} F_{\alpha i}) + \frac{1}{2}(\delta F_{\alpha i} F_{\alpha j,k} + \delta F_{\alpha j} F_{\alpha i,k}). \end{aligned} \right\} \qquad (2.22)$$

The following relations between the Eulerian stress and hyperstress tensors and the Piola–Kirchhoff tensors $S_{ij}$ and $P_{ijk}$ are found to be:

$$\left. \begin{aligned} \varsigma_\alpha &= 0, \qquad \Sigma_{\alpha\beta} = J^{-1}[S_{ij} F_{\alpha i} F_{\beta j} + P_{ijk}(F_{\alpha j} F_{\beta i,k} + F_{\alpha i,k} F_{\beta j})], \\ \Pi_{\alpha\beta\gamma} &= J^{-1} P_{ijk} F_{\alpha j} F_{\beta i} F_{\gamma k}. \end{aligned} \right\} \qquad (2.23)$$

Furthermore, bearing in mind equation (2.16), we assume the existence of an objective stored energy function $\tilde{\psi}(E_{ik}, E_{ik,j})$, the integral of which gives the global stored energy of the body. We also assume that the variation of the global stored deformation energy equals the internal work (2.21)

$$\delta\left[\int_{\mathcal{D}_0} \tilde{\psi}(E_{ik}, E_{ik,j})\right] = -\tilde{W}^{\text{int}}. \qquad (2.24)$$

This assumption yields the following expressions:

$$s_i = 0, \qquad S_{ij} = \frac{\partial \tilde{\psi}}{\partial E_{ij}}, \qquad P_{ijk} = \frac{\partial \tilde{\psi}}{\partial E_{ij,k}}, \qquad (2.25)$$

for the zero-order stress field $s$, the second Piola–Kirchhoff stress tensor $S$ and the referential hyperstress tensor $P$. We remark that $P$ and $\Pi$ are symmetric with respect to their first two indices, and we recall that the contribution of $\Pi$ to the Eulerian internal work is $\Pi_{\alpha\beta\gamma}u_{\alpha,\beta\gamma}$, being $u_{\alpha,\beta\gamma}$ symmetric with respect to its last two indices. However, using the decomposition (2.10) for $\Pi$ it is easily proved that both the completely symmetric and the sym-skew parts of $\Pi$ contribute to the internal work.

The local equilibrium equations are deduced by requiring the validity of equation (2.2) for every kinematically admissible virtual displacement field. To this aim, the distinction between essential (prescribed value of displacement) and natural (prescribed value of traction) boundary conditions needs to be generalized. On every part $\mathcal{L}_x^y \subset \partial \mathcal{D}$, with $x=e,n$, $y=e,n$, we may impose four different kinds of boundary conditions; more precisely we define:

| | | | | |
|---|---|---|---|---|
| $\mathcal{L}_e^e$ | $u_\alpha = \bar{u}_\alpha$ | $u_{\alpha,\beta}m_\beta = \bar{w}_\alpha$ | $u$-essential | $(D_n u)$-essential |
| $\mathcal{L}_e^n$ | $u_\alpha = \bar{u}_\alpha$ | $\tau_\alpha = \bar{\tau}_\alpha$ | $u$-essential | $\tau$-natural |
| $\mathcal{L}_n^e$ | $t_\alpha = \bar{t}_\alpha$ | $u_{\alpha,\beta}m_\beta = \bar{w}_\alpha$ | $t$-natural | $(D_n u)$-essential |
| $\mathcal{L}_n^n$ | $t_\alpha = \bar{t}_\alpha$ | $\tau_\alpha = \bar{\tau}_\alpha$ | $t$-natural | $\tau$-natural |

where $(D_n u)_\alpha := u_{\alpha,\beta} m_\beta$ indicates the normal derivative of $u$ and $m$ is the outward normal to $\partial\mathcal{D}$. Subscripts distinguish $u$-essential from $t$-natural boundary conditions, while superscripts distinguish $(D_n u)$-essential from $\tau$-natural boundary conditions; moreover, on every part of the edge $\mathcal{E}_h$ standard distinction between essential and natural conditions holds true.

The varied displacement $u^*$ is said to be kinematically admissible if it satisfies the same essential boundary conditions as $u$; as a consequence of equation (2.4) the virtual displacement $\delta u$ and its normal derivative, both to be used in the principle of virtual working, will vanish on $\partial\mathcal{D}_e^\star := \mathcal{L}_e^e \cup \mathcal{L}_e^n$ and $\partial\mathcal{D}_\star^e := \mathcal{L}_e^e \cup \mathcal{L}_n^e$, respectively.

Integration by parts of the left-hand side of (2.24) and standard localization arguments yield

$$\left.\begin{aligned}
[F_{\alpha i}(S_{ij} - P_{ijk,k})]_{,j} + Jb_\alpha &= 0, & &\text{on } \mathcal{D}_0, \\
F_{\alpha i}(S_{ij} - P_{ijk,k})n_j - (Q_{Bj}F_{\alpha i}P_{ijk}n_k)_{,B} &= J_S t_\alpha, & &\text{on } (\partial\mathcal{D}_0)_n^\star, \\
F_{\alpha i}P_{ijk}n_k n_j &= J_S \tau_\alpha, & &\text{on } (\partial\mathcal{D}_0)_\star^n, \\
[\![Q_{Bj}F_{\alpha i}P_{ijk}n_k \nu_B]\!] &= J_L f_\alpha, & &\text{on } \mathcal{E}_{0h}^n,
\end{aligned}\right\} \quad (2.26)$$

which must be used together with the essential boundary conditions

$$u_\alpha = \bar{u}_\alpha \text{ on } (\partial\mathcal{D}_0)_e^\star, \qquad u_{\alpha,\beta}n_\beta = \bar{w}_\alpha \text{ on } (\partial\mathcal{D}_0)_\star^e, \qquad u_\alpha = \bar{u}_\alpha \text{ on } \mathcal{E}_{0h}^e. \quad (2.27)$$

Here,

— $n$ denotes the outward normal to $\partial\mathcal{D}_0$;
— every part of the boundary $\partial\mathcal{D}_0$ is locally parametrized by the coordinate system $\hat{X}_B$ for $B=1,2$; by definition $Q_{Bi} := \partial \hat{X}_B / \partial X_i$ and $(\ )_{,B}$ denotes derivation with respect to the $\hat{X}_B$ coordinate;
— as $\partial\mathcal{D}_0$ and $\mathcal{E}_0$ are the inverse images of $\partial\mathcal{D}$ and $\mathcal{E}$ under the placement map $\chi$, thus $(\partial\mathcal{D}_0)_e^\star$ is the inverse image of $\partial\mathcal{D}_e^\star$ under $\chi$ and $(\partial\mathcal{D}_0)_\star^e$ is that of $\partial\mathcal{D}_\star^e$;
— $J_S$, $J_L$ are the determinant of the surface and line restriction of $F$;

— every edge of $\partial\mathcal{D}_0$ corresponds to a jump of its outward normal unit vector. Let $\mu_i$ denote the components of the unit tangent vector to an edge and $n_i^-, n_i^+$ be the two values of $n$ on the two faces of the edge itself; then on the same edge $\nu_i^\pm := \epsilon_{ijl}\mu_j n_l^\pm$ indicate the components of the Darboux tangent-normal vectors on the left and the right-hand side of the edge; $(\mu_i, \nu_i^+, n_i^+)$ and $(\mu_i, \nu_i^-, n_i^-)$ form the two left-hand bases characterizing both sides of the discontinuity; and

— $[\![\varphi]\!] := (\varphi)^+ - (\varphi)^-$ indicates the jump through the edge.

The last term in the left-hand side of equation $(2.26)_2$ can be found also in Mindlin (1964) and Germain (1973); it arises when splitting a divergence term into a divergence on the surface plus a normal derivative term, namely $(-P_{ijk,k}n_j)$, which is naturally added to the standard Cauchy traction, $S_{ij}n_j$.

## 3. Linear isotropic constitutive relations

Generalization of Hooke's law to second gradient materials stems from requiring the constitutive relation among generalized stresses ($S$ and $P$) and strains ($E$ and $\nabla E$) to be linear and isotropic. Linearity of the constitutive relations (2.25) with respect to the strain measures is enforced, as usual, assuming that the stored elastic energy $\tilde{\psi}$ is a quadratic form of both its arguments

$$\tilde{\psi}(E_{ij}, E_{ij,k}) = \frac{1}{2}(C_{ijkl}E_{ij}E_{kl} + 2H_{ijklp}E_{ij,k}E_{lp} + G_{ijklpq}E_{ij,k}E_{lp,q}), \qquad (3.1)$$

where the fourth-order tensor $C$, the fifth-order tensor $H$, and the sixth-order tensor $G$ possess, without loss of generality, the symmetries

$$C_{ijkl} = C_{klij}, \qquad H_{ijklp} = H_{lpijk}, \qquad G_{ijklpq} = G_{lpqijk}. \qquad (3.2)$$

Moreover, the symmetry of the Green–Lagrange strain induces the following additional symmetries:

$$C_{ijkl} = C_{ijlk} = C_{jikl}, \qquad H_{ijklp} = H_{jiklp} = H_{ijkpl}, \qquad G_{ijklpq} = G_{jiklpq} = G_{ijkplq}. \qquad (3.3)$$

When (3.1) holds, the constitutive relations (2.25) read as follows:

$$S_{ij} = C_{ijkl}E_{kl} + H_{klpij}E_{kl,p}, \qquad P_{ijk} = H_{ijklp}E_{lp} + G_{ijklpq}E_{lp,q}. \qquad (3.4)$$

Following standard arguments for material symmetry characterization, a second gradient material is said to be isotropic if the stored elastic energy satisfies the following:

$$\tilde{\psi}(E_{ij}, E_{ij,k}) = \tilde{\psi}(Q_{hi}E_{hm}Q_{mj}, Q_{hi}E_{hm,n}Q_{mj}Q_{nk}), \qquad (3.5)$$

for every orthogonal transformation $Q \in$ Orth. The request (3.5) implies the elasticity tensors to fulfil,

$$\left.\begin{aligned}C_{ijkl} &= C_{hmnr}Q_{hi}Q_{mj}Q_{nk}Q_{rl}, \\ H_{ijklp} &= H_{hmnrs}Q_{hi}Q_{mj}Q_{nk}Q_{rl}Q_{sp}, \\ G_{ijklpq} &= G_{hmnrst}Q_{hi}Q_{mj}Q_{nk}Q_{rl}Q_{sp}Q_{tq},\end{aligned}\right\} \qquad (3.6)$$

for every $Q \in \text{Orth}$. In Suiker & Chang (2000) the tensors $C$, $H$ and $G$ satisfying equations (3.6) are determined considering only proper orthogonal transformations $Q \in \text{Orth}^+$. However, this is not sufficient to characterize isotropy; indeed, equations (3.6) must also be fulfilled by every reflection.[2]

This distinction is not relevant when discussing isotropy of even-order tensors, $(3.6)_{1,3}$, but reveals to be crucial when considering odd-order tensors as in $(3.6)_2$.

Requiring the conditions (3.2), (3.3) and (3.6) to be satisfied for every $Q \in \text{Orth}$, we obtain generalized Hooke's law for isotropic second gradient materials

$$\left.\begin{aligned}
C_{ijkl} &= \lambda \delta_{ij}\delta_{kl} + \mu(\delta_{ik}\delta_{jl} + \delta_{il}\delta_{jk}), \qquad H_{ijklp} = 0, \\
G_{ijklpq} &= c_2(\delta_{ij}\delta_{kl}\delta_{pq} + \delta_{ij}\delta_{kp}\delta_{lq} + \delta_{ik}\delta_{jq}\delta_{lp} + \delta_{iq}\delta_{jk}\delta_{lp}) + c_3(\delta_{ij}\delta_{kq}\delta_{lp}) \\
&+ c_5(\delta_{ik}\delta_{jl}\delta_{pq} + \delta_{ik}\delta_{jp}\delta_{lq} + \delta_{il}\delta_{jk}\delta_{pq} + \delta_{ip}\delta_{jk}\delta_{lq}) + c_{11}(\delta_{il}\delta_{jp}\delta_{kq} \\
&+ \delta_{ip}\delta_{jl}\delta_{kq}) + c_{15}(\delta_{il}\delta_{jq}\delta_{kp} + \delta_{ip}\delta_{jq}\delta_{kl} + \delta_{iq}\delta_{jl}\delta_{kp} + \delta_{iq}\delta_{jp}\delta_{kl}).
\end{aligned}\right\} \quad (3.7)$$

Therefore, isotropic linear elastic second gradient materials are completely described in terms of seven constants; two of them are the standard Lamé coefficients $\lambda$ and $\mu$. In (3.7) a notation similar to that introduced in Suiker & Chang (2000) was used for the material constants. Remark that

(i) in Suiker & Chang (2000) the equations (3.7) are not explicitly derived; as long as one needs to consider the boundary conditions $(2.26)_{2,3,4}$, they must be derived;
(ii) without the request of invariance under reflections the fifth-order tensor $H$ would not vanish, being in the form

$$H_{ijklp} = c_8(\varepsilon_{ikl}\delta_{jp} + \varepsilon_{ikp}\delta_{jl} + \varepsilon_{jkl}\delta_{ip} + \varepsilon_{jkp}\delta_{il}), \quad (3.8)$$

which is the form of 'isotropic' fifth-order tensor determined by Suiker & Chang (2000), where isotropy is defined as invariance under linear transformations in $\text{Orth}^+$.

## 4. Positive definiteness of stored elastic energy

As well known in linear elasticity,[3] if the stored elastic energy $\psi$ is a strictly convex function of strain, then the solution of the equilibrium equations, in terms of displacements, is unique up to an infinitesimal rigid displacement field (see Salençon 1995). A first approach to second-gradient elasticity may start by assuming that the stored elastic energy is a strictly convex function of strain and strain-gradient. Investigations will be needed to generalize quasi-convexity and related arguments.

Assuming $\psi$ to be a quadratic form of its arguments, the strict convexity is equivalent to the positive definiteness of the following quadratic form of $E$ and $\nabla E$:

$$\psi = \frac{1}{2} S_{ij}(E_{lm}, E_{lm,p}) E_{ij} + P_{ijk}(E_{lm}, E_{lm,p}) E_{ij,k} > 0,$$

---

[2] Every $Q \in \text{Orth}$ can be obtained as the composition of an element of $\text{Orth}^+$ either with the identity matrix ($I$) or with an arbitrary reflection; for instance minus the identity ($-I$) when considering an odd dimensional ambient space.

[3] Here linearity means that only the linear part of the Green–Lagrange strain tensor with respect to displacement gradient is retained.

where $S$ and $P$ are prescribed by equation (3.4). In the case of isotropic materials, see equation (3.7), the positive definiteness of $\psi$ yields inequality constraints on both first and second gradient constitutive parameters $\lambda$, $\mu$, $c_2$, $c_3$, $c_5$, $c_{11}$ and $c_{15}$.

In order to determine these constraints, one can rewrite equation (3.7) in matrix form, using a Voigt-type representation for the considered tensors. The Sylvester criterion[4] is then applied to check the positive definiteness of the quadratic form $\psi$. To this aim, different equivalent choices for listing the strict components of tensors $E$, $\nabla E$, $S$ and $P$ can be introduced.

As the constitutive coupling $H$ vanishes, the standard Voigt representation for $S$ and $E$ allows for the independent determination of the constraints $\mu > 0$ and $3\lambda + 2\mu > 0$. Then, considering only the relationship between $P$ and $\nabla E$, the equations (3.7) can be stated in the following block-diagonal form:

$$\left.\begin{array}{c}\begin{pmatrix}P_{111}\\P_{122}\\P_{133}\\P_{221}\\P_{331}\end{pmatrix}=G_1\begin{pmatrix}E_{11,1}\\E_{12,2}\\E_{13,3}\\E_{22,1}\\E_{33,1}\end{pmatrix},\quad\begin{pmatrix}P_{222}\\P_{121}\\P_{233}\\P_{112}\\P_{332}\end{pmatrix}=G_1\begin{pmatrix}E_{22,2}\\E_{12,1}\\E_{23,3}\\E_{11,2}\\E_{33,2}\end{pmatrix},\\ \begin{pmatrix}P_{333}\\P_{131}\\P_{232}\\P_{113}\\P_{223}\end{pmatrix}=G_1\begin{pmatrix}E_{33,3}\\E_{13,1}\\E_{23,2}\\E_{11,3}\\E_{22,3}\end{pmatrix},\quad\begin{pmatrix}P_{123}\\P_{132}\\P_{231}\end{pmatrix}=G_2\begin{pmatrix}E_{12,3}\\E_{13,2}\\E_{23,1}\end{pmatrix},\end{array}\right\} \quad (4.1)$$

where $G_1$ and $G_2$ matrices are defined as follows:

$$G_1 := \begin{pmatrix} 4c_2+c_3+4c_5+2c_{11}+4c_{15} & 2c_2+4c_5 & 2c_2+4c_5 & 2c_2+c_3 & 2c_2+c_3 \\ 2c_2+4c_5 & 4(c_5+c_{11}+c_{15}) & 4c_5 & 2c_2+4c_{15} & 2c_2 \\ 2c_2+4c_5 & 4c_5 & 4(c_5+c_{11}+c_{15}) & 2c_2 & 2c_2+4c_{15} \\ 2c_2+c_3 & 2c_2+4c_{15} & 2c_2 & c_3+2c_{11} & c_3 \\ 2c_2+c_3 & 2c_2 & 2c_2+4c_{15} & c_3 & c_3+2c_{11} \end{pmatrix}$$

and

$$G_2 := 4\begin{pmatrix} c_{11} & c_{15} & c_{15} \\ c_{15} & c_{11} & c_{15} \\ c_{15} & c_{15} & c_{11} \end{pmatrix}.$$

However, to apply Sylvester's criterion to matrix $G_1$ may present some difficulties. To overcome them, the decomposition (2.10) for the tensors $P$ and $\nabla E$ is revealed to be useful. Indeed, once a suitable linear combination of the strict components of $P$ and $\nabla E$ is introduced, the constitutive relations (3.7) assume the form

---

[4] Sylvester's criterion states that a matrix $M$ is positive definite iff the determinants associated with all upper-left submatrices of $M$ are positive.

$$\begin{pmatrix} \tilde{P}_{111} \\ \tilde{P}_{122} + \tilde{P}_{133} \\ \hat{P}_{32} - \hat{P}_{23} \end{pmatrix} = \Gamma_1 \begin{pmatrix} \tilde{E}_{111} \\ \tilde{E}_{122} + \tilde{E}_{133} \\ \hat{E}_{32} - \hat{E}_{23} \end{pmatrix},$$

$$\begin{pmatrix} \tilde{P}_{222} \\ \tilde{P}_{233} + \tilde{P}_{112} \\ \hat{P}_{13} - \hat{P}_{31} \end{pmatrix} = \Gamma_1 \begin{pmatrix} \tilde{E}_{222} \\ \tilde{E}_{233} + \tilde{E}_{112} \\ \hat{E}_{13} - \hat{E}_{31} \end{pmatrix},$$

$$\begin{pmatrix} \tilde{P}_{333} \\ \tilde{P}_{223} + \tilde{P}_{113} \\ \hat{P}_{21} - \hat{P}_{12} \end{pmatrix} = \Gamma_1 \begin{pmatrix} \tilde{E}_{333} \\ \tilde{E}_{223} + \tilde{E}_{113} \\ \hat{E}_{21} - \hat{E}_{12} \end{pmatrix},$$

(4.2)

$$\left.\begin{aligned} \tilde{P}_{122} - \tilde{P}_{133} &= 3\gamma_2(\tilde{E}_{122} - \tilde{E}_{133}), & \hat{P}_{32} + \hat{P}_{23} &= 6\gamma_5(\hat{E}_{32} + \hat{E}_{23}), \\ \tilde{P}_{233} - \tilde{P}_{112} &= 3\gamma_2(\tilde{E}_{233} - \tilde{E}_{112}), & \hat{P}_{13} + \hat{P}_{31} &= 6\gamma_5(\hat{E}_{13} + \hat{E}_{31}), \\ \tilde{P}_{223} - \tilde{P}_{113} &= 3\gamma_2(\tilde{E}_{223} - \tilde{E}_{113}), & \hat{P}_{21} + \hat{P}_{12} &= 6\gamma_5(\hat{E}_{21} + \hat{E}_{12}), \end{aligned}\right\}$$

(4.3)

$$\begin{pmatrix} \hat{P}_{11} \\ \hat{P}_{22} \\ \hat{P}_{33} \end{pmatrix} = \Gamma_2 \begin{pmatrix} \hat{E}_{11} \\ \hat{E}_{22} \\ \hat{E}_{33} \end{pmatrix}, \qquad \tilde{P}_{123} = 3\gamma_2 \tilde{E}_{123},$$

(4.4)

where

$$\Gamma_1 := \begin{pmatrix} \gamma_1 & 2\gamma_1 - \gamma_2 & \gamma_3 \\ 2\gamma_1 - \gamma_2 & 4\gamma_1 + \gamma_2 & 2\gamma_3 \\ \gamma_3 & 2\gamma_3 & \gamma_4 \end{pmatrix}, \qquad \Gamma_2 := \gamma_5 \begin{pmatrix} 2 & -1 & -1 \\ -1 & 2 & -1 \\ -1 & -1 & 2 \end{pmatrix}$$

(4.5)

and

$$\left.\begin{aligned} \gamma_1 &:= 2(c_{11} + 2c_{15}) + 4c_2 + c_3 + 4c_5, \\ \gamma_2 &:= 4(c_{11} + 2c_{15}), \gamma_3 := \frac{2}{3}(4c_5 - 2c_2 - 2c_3), \\ \gamma_4 &:= \frac{8}{9}(3c_{11} - 3c_{15} - 4c_2 + 2c_3 + 2c_5), \gamma_5 := \frac{4}{9}(c_{11} - c_{15}). \end{aligned}\right\}$$

(4.6)

The use of the introduced decomposition reduces to three the maximal dimension of the constitutively coupled blocks and renders the application of the Sylvester criterion feasible. Indeed, positive definiteness of $\psi$ implies the following conditions on the constitutive parameters $\gamma_i$:

$$\gamma_1 > 0, \qquad 0 < \gamma_2 < 5\gamma_1, \qquad \gamma_4 > \frac{5\gamma_3^2}{5\gamma_1 - \gamma_2}, \qquad \gamma_5 > 0, \qquad (4.7)$$

which can be also stated in terms of the constitutive parameters $c_i$

$$\left.\begin{array}{c} c_{11} > 0, \qquad -\dfrac{c_{11}}{2} < c_{15} < c_{11}, \qquad 5c_3 + 4c_{11} > 2c_{15}, \\[6pt] c_5 > \dfrac{c_3(3c_{11} + c_{15}) + 2\left(c_{11}^2 - 5c_2^2 - 6c_{15}c_2 - 2c_{15}^2 + c_{11}(2c_2 + c_{15})\right)}{4c_{15} - 10c_3 - 8c_{11}}. \end{array}\right\} \quad (4.8)$$

The decomposition proposed for the strict components of the hyperstress $P$ and strain gradient $\nabla E$ is illustrated in appendix A.

Under the assumption of small strains, Sokolowski (1970), closely following Mindlin (1964), assumes the stored energy to depend quadratically on the infinitesimal strain $\epsilon_{ij} := (u_{i,j} + u_{j,i})/2$ and on the gradient of the curl of the displacement field $\kappa_{ij} := \omega_{i,j} = -\varepsilon_{ipq} u_{p,qj}$. Two additional constants, $\eta$ and $\ell$, are hence introduced in his constitutive relations. We remark that, the second-order tensor $\kappa_{ij}$, which is the only second gradient contribution considered by Toupin (1962), Mindlin (1964) and Sokolowski (1970), coincides with the linear part of $\hat{E}_{ij}$ with respect to $\xi := \|\nabla u\| \ll 1$:

$$\kappa_{ij} := \omega_{i,j} = \dfrac{\partial \hat{E}_{ij}}{\partial \xi}\Big|_{\xi=0}. \tag{4.9}$$

However, equations (4.2) and (4.5) show that, even for linear isotropic materials, the completely symmetric strain gradient $\tilde{E}$ is constitutively coupled with the sym-skew strain gradient $\hat{E}$.

The constitutive relations (4.1), or equivalently (4.2)–(4.4), can be compared with those of Sokolowski (1970) when assuming small deformations around a zero-stress reference configuration. If the five constitutive second gradient constants are chosen according to the following:

$$\left.\begin{array}{c} c_2 = \ell^2 \eta \mu, \qquad c_3 = -2\ell^2 \eta \mu, \qquad c_5 = -\ell^2 \eta \mu/2, \\[4pt] c_{11} = \ell^2 (\eta + 1)\mu, \qquad c_{15} = -\ell^2 (\eta + 1)\mu/2 \end{array}\right\} \quad (4.10)$$

or

$$\gamma_1 = \gamma_2 = \gamma_3 = 0, \qquad \gamma_4 = 4\ell^2 (1-\eta)\mu, \qquad \gamma_5 = \dfrac{2}{3}\ell^2 (1+\eta)\mu, \tag{4.11}$$

our constitutive relations are equivalent to the ones derived by Sokolowski (1970). For the particular case (4.10), the conditions (4.8) reduce to

$$\ell^2 > 0, \qquad -1 < \eta < 1, \tag{4.12}$$

as required in Sokolowski (1970).

## 5. Warping in torsion of a prismatic second gradient cylinder

This section is devoted to show how second gradient theories modify the elliptic problem for warping in the torsion of prismatic cylinders. To this aim, the linearized form of strain measures (2.5) in terms of displacement gradient, and the balance laws (2.26) near a zero-stress reference configuration will be considered. In particular the infinitesimal strain $\epsilon_{ij}$ is used instead of the Green–Lagrange strain $E_{ij}$, while the Piola–Kirchhoff and the Cauchy stress (and hyperstress) can be identified. Let us introduce an orthonormal coordinate system $\{X_1, X_2, X_3\}$ for the reference configuration of a prismatic cylinder, in which the $X_3$ axis is orthogonal to the cross section $\mathcal{B}$. The body forces in the cylinder and the tractions on the mantle are supposed vanishing as in the standard *de Saint Venant* problem; forces can be applied only on the two bases and on their boundaries. As displacement field for such a cylinder we here postulate the standard displacement field of the *de Saint Venant* torsion,

$$u_1 = -\Theta X_2 X_3, \qquad u_2 = \Theta X_1 X_3, \qquad u_3 = \Theta w(X_1, X_2), \tag{5.1}$$

where $\Theta$ is the unit angle of torsion. Accordingly, in what follows, we deduce the balance laws, that the unknown function $w(X_1, X_2)$ must satisfy, and the boundary conditions, that one must apply to the bases, for the field (5.1) to be a solution. It is easy to verify from (5.1) that the only non-vanishing components of strain and strain gradient are the following:

$$\left.\begin{aligned}
\varepsilon_{13} &= \varepsilon_{31} = \Theta(w_{,1} - X_2)/2, & \varepsilon_{23} &= \varepsilon_{32} = \Theta(w_{,2} + X_1)/2, \\
\varepsilon_{13,2} &= \varepsilon_{31,2} = \Theta(w_{,12} - 1)/2, & \varepsilon_{23,1} &= \varepsilon_{32,1} = \Theta(w_{,12} + 1)/2, \\
\varepsilon_{13,1} &= \varepsilon_{31,1} = \Theta w_{,11}/2, & \varepsilon_{32,2} &= \varepsilon_{23,2} = \Theta w_{,22}/2.
\end{aligned}\right\} \tag{5.2}$$

Using the constitutive relation (3.7), we obtain the corresponding non-vanishing stresses and hyperstresses

$$\left.\begin{aligned}
S_{13} &= S_{31} = \mu\Theta(w_{,1} - X_2)/2, \quad S_{23} = S_{32} = \mu\Theta(w_{,2} + X_1)/2, \\
P_{113} &= \Theta(2c_{15} w_{,11} + c_2 \Delta w), \quad P_{123} = P_{213} = \Theta(2c_{15} w_{,12}), \\
P_{131} &= P_{311} = \Theta((c_{11} + c_{15}) w_{,11} + c_5 \Delta w), \\
P_{132} &= P_{312} = \Theta(c_{11}(w_{,12} - 1) + c_{15}(w_{,12} + 1)), \\
P_{223} &= \Theta(2c_{15} w_{,22} + c_2 \Delta w), \\
P_{231} &= P_{321} = \Theta(c_{15}(w_{,12} - 1) + c_{11}(w_{,12} + 1)), \\
P_{232} &= P_{322} = \Theta((c_{11} + c_{15}) w_{,22} + c_5 \Delta w), \\
P_{333} &= \Theta(c_2 + 2c_5) \Delta w,
\end{aligned}\right\} \tag{5.3}$$

where $\Delta w := w_{,11} + w_{,22}$ means the in-plane Laplacian.

Finally, substituting in the linearized form of the balance laws (2.26), we obtain the following elliptic problem for the warping function $w$:

$$\begin{cases}
\mu \Delta w - (c_{11} + c_{15} + c_5) \Delta \Delta w = 0, & \text{in } \mathcal{B}, \\
(\mu w - (c_{11} + c_{15} + c_5) \Delta w)_{,B} n_B + d_S = \mu \varepsilon_{AB} X_B n_A, & \text{on } \partial \mathcal{B}, \\
N_{AB} w_{,AB} = 0, & \text{on } \partial \mathcal{B},
\end{cases} \tag{5.4}$$

where

$$d_S := \frac{c_{11} + c_{15}}{2} \frac{\partial}{\partial s} \left( \sin 2\vartheta (w_{,22} - w_{,11}) + 2 \cos 2\vartheta\, w_{,12} \right) \tag{5.5}$$

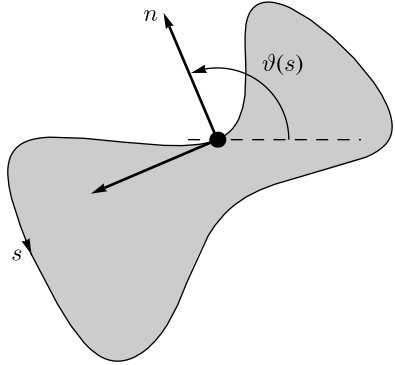

Figure 1. The conventions used for the abscissa $s$ and the angle $\vartheta$ to compute the divergence $d_S$ on the mantle of the cylinder.

and

$$\epsilon_{AB} := \begin{pmatrix} 0 & -1 \\ 1 & 0 \end{pmatrix}_{AB}, \qquad N_{AB} := (c_{11} + c_{15})n_A n_B + c_5 \delta_{AB}. \qquad (5.6)$$

Here, $\Delta\Delta$ means the in-plane double Laplacian, $A,B=1,2$ and summation over repeated indices is understood. The abscissa $s$ along the boundary $\partial \mathcal{B}$ and the angle $\vartheta$ formed between the normal $n$ and the horizontal axis have been chosen according to figure 1. Let us remark that the term $d_S$ introduces in the boundary conditions $(5.4)_4$ the curvature $\partial \vartheta/\partial s$ of the curve $\partial \mathcal{B}$, as one would expect when considering the general form of the boundary conditions $(2.26)_2$ for the traction.

The obtained elliptic problem does not coincide with the one formulated in Sokolowski (1970), because the term corresponding there to $d_S$ does not contain the curvature of the boundary $\partial \mathcal{B}$. Moreover, in Sokolowski (1970) the matrix $N_{AB}$ is spherical and the boundary condition $(5.4)_3$ reduces to $\partial^2 w/\partial n^2 = 0$. This last circumstance is also related to the more general constitutive equations (4.1) found in §3. Indeed, Sokolowski (1970) excludes from the analysis the completely symmetric part of the strain gradient, hence limiting to the particular class of second gradient materials in which hyperstresses reduce to the so called couple-stress.

In the case of a circular hollow cylinder, the elliptic problem (5.4) has $w=0$ as unique solution and the torsion problem can be easily solved. In this case, the deformation fields (5.2) as well as the stress fields (5.3) are sensibly simplified. Indeed, when $w=0$, the displacement field (5.1) depends quadratically on the variables $\{X_i\}$, the hyperstress components $P_{ijk}$ result to be constant and the local equilibrium equations $(2.26)_1$ in the bulk are identically verified. Moreover, the boundary conditions $(2.26)_{2,3}$ are verified with vanishing tractions $t$ and double forces $\tau$ on the lateral mantle of the cylinder. On the other hand, the tractions and double forces $(2.26)_{2,3}$ on the basis $\mathcal{B}_L$ at $X_3=L$, as well as the concentrated edge forces $(2.26)_4$ on the inner and outer circumferences $\mathcal{C}_L^\pm$, do not vanish and are estimated as follows:

$$\begin{cases} t_1 = -\mu \Theta X_2, \quad t_2 = \mu \Theta X_1, \quad t_3 = 0, \quad \text{in } \mathcal{B}_L, \\ \tau_\alpha = 0, \alpha = 1,2,3 \quad \text{in } \mathcal{B}_L, \\ f_1 = \Theta(c_{11} - c_{15})\sin\vartheta, \quad f_2 = -\Theta(c_{11} - c_{15})\cos\vartheta, \quad f_3 = 0, \quad \text{on } \mathcal{C}_L^\pm. \end{cases} \qquad (5.7)$$

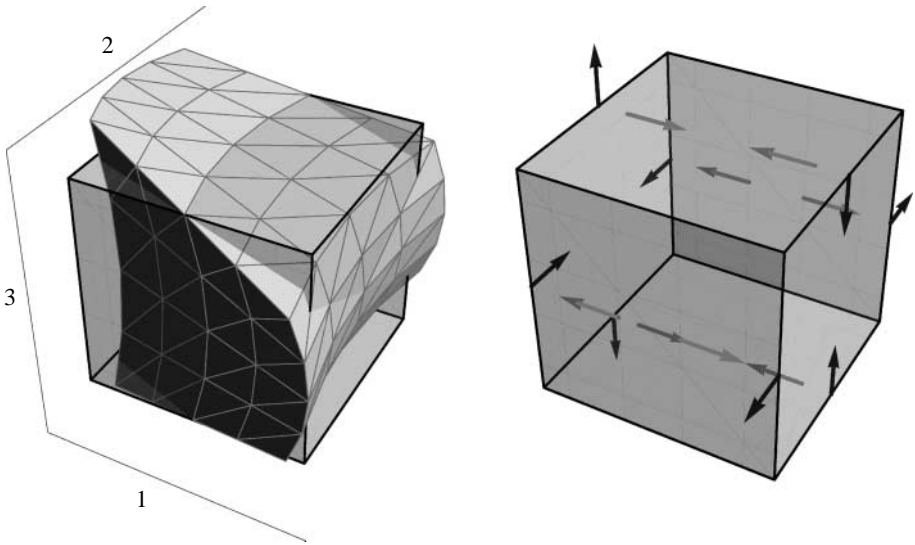

Figure 2. The second gradient deformation and hyperstress components: $\tilde{E}_{122}-\tilde{E}_{133}$ and $\tilde{P}_{122}-\tilde{P}_{133}$. These are constitutively coupled by $\gamma_2$ in equation $(4.3)_1$.

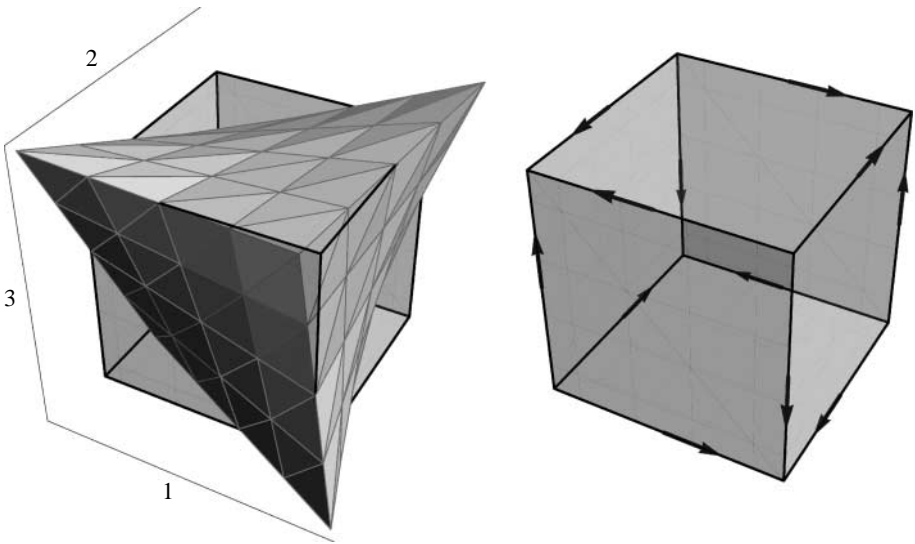

Figure 3. The second gradient deformation and hyperstress components: $\tilde{E}_{123}$ and $\tilde{P}_{123}$. These are coupled by $\gamma_2$ in equation $(4.4)_2$.

For the circular hollow cross section, the angle $\vartheta$ coincides with the circumferential polar coordinate in the basis. The surface tractions, the surface double-forces and the tractions per unit line given in (5.7) are statically equivalent to a pure torque.

The simplified form of (5.2) and (5.3) for $w=0$, allows for the closed form evaluation of the torsional stiffness, say $K_t$; to this aim, the stored elastic energy is written as the following quadratic function of $\Theta$:

$$\psi = \frac{1}{2}\int_{\mathcal{D}}(S_{ij}\epsilon_{ij} + P_{ijk}\epsilon_{ij,k}) =: \frac{1}{2}K_t\Theta^2. \qquad (5.8)$$

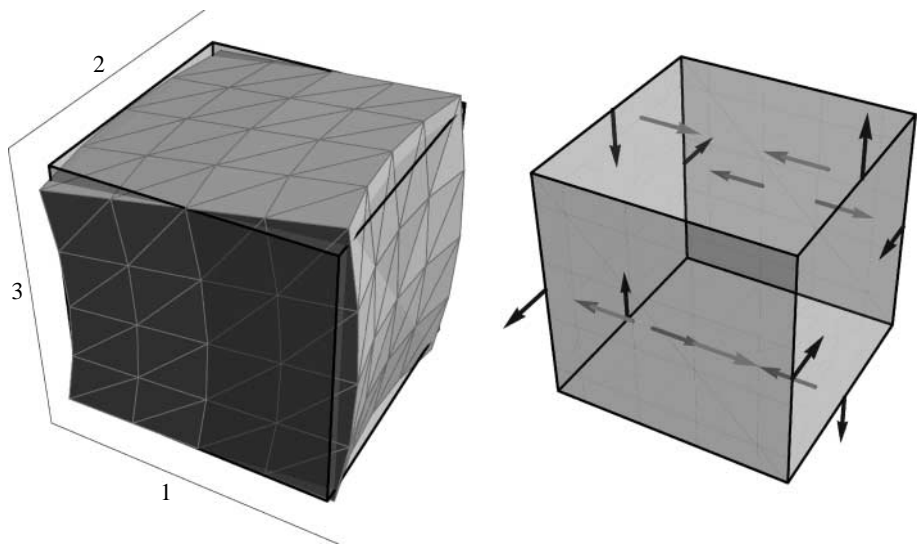

Figure 4. The second gradient deformation and hyperstress components: $\hat{E}_{32} + \hat{E}_{23}$ and $\hat{P}_{32} + \hat{P}_{23}$. These are coupled by $\gamma_4$ in equation $(4.3)_2$.

With simple calculations one obtains the following:
$$K_t = \mu I_P + 2(c_{11} - c_{15})A, \tag{5.9}$$
where $I_P := \pi(R_{\text{ext}}^4 - R_{\text{int}}^4)/2$ is the moment of polar inertia, $A := \pi(R_{\text{ext}}^2 - R_{\text{int}}^2)$ is the area of the cross section, $R_{\text{int}}$ and $R_{\text{ext}}$ are the inner and outer radii of the cylinder. Equation (5.9) may represent an effective tool to experimentally estimate the second gradient constitutive parameter $\gamma_5 = 4(c_{11} - c_{15})/9$. Moreover, the analytical solution found could be a valid (patch) test in the development of finite-element codes for second gradient materials. From an experimental point of view, the solution with $w = 0$ can be realized clamping the two bases of a hollow circular cylinder to two rigid supports which undergo a relative rotation $\Theta$ around the $X_3$ axis.

## 6. Conclusions

A geometrically nonlinear theory of second gradient elasticity has been analysed; this accounts for Green–Lagrange strain and strain-gradient measures as well as for the Cauchy and Piola–Kirchhoff stress and hyperstress tensors. Owing to a previous result (Fortuné & Vallée 2001) which links the rotation gradient to the strain gradient, the most general hyperelastic energy functional must depend on the strain and on all the components of strain gradient including the completely symmetric part. Hence, hyperstresses do not limit to couple stresses but include double forces.

The most general linear elastic constitutive relations for isotropic materials are also derived; these relations generalize standard Hooke's law to second gradient materials. The presented analysis shows that a complete isotropic second gradient constitutive theory must include five more parameters in addition to the classical Lamé constants. The conditions for the positive definiteness of the considered quadratic strain energy are found by means of a novel decomposition of the strain gradient tensor, graphically shown in appendix A.

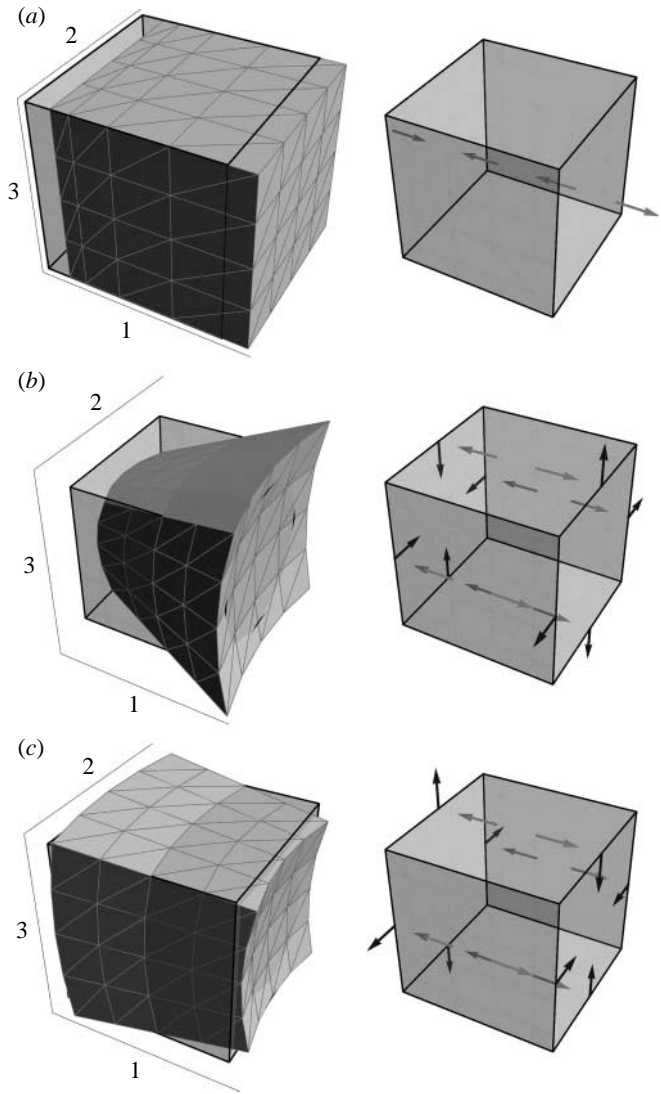

Figure 5. The second gradient deformation and hyperstress components: (a) $\tilde{E}_{111}$ and $\tilde{P}_{111}$; (b) $\tilde{E}_{122} + \tilde{E}_{133}$ and $\tilde{P}_{122} + \tilde{P}_{133}$; and (c) $\hat{E}_{32} - \hat{E}_{23}$ and $\hat{P}_{32} - \hat{P}_{23}$. These are coupled by $\Gamma_1$ in equation $(4.2)_1$.

Even in the simple problem of linear torsion formulated in §5, the introduced constitutive parameters play a relevant role, influencing the warping function and the torsional stiffness. While the experimental measurement of the second gradient parameter $\gamma_5$ seems feasible, from the quoted results, it is an open problem the design of measurement procedures for the remaining four parameters. For instance, the correction to the bending stiffness found by Anthoine (1998) under the constitutive assumption of Sokolowski (1970), could be extended to consider the general constitutive relations proposed here.

The authors thank dott. Giuseppe Ruta for having discussed carefully the content of a first draft of this paper.

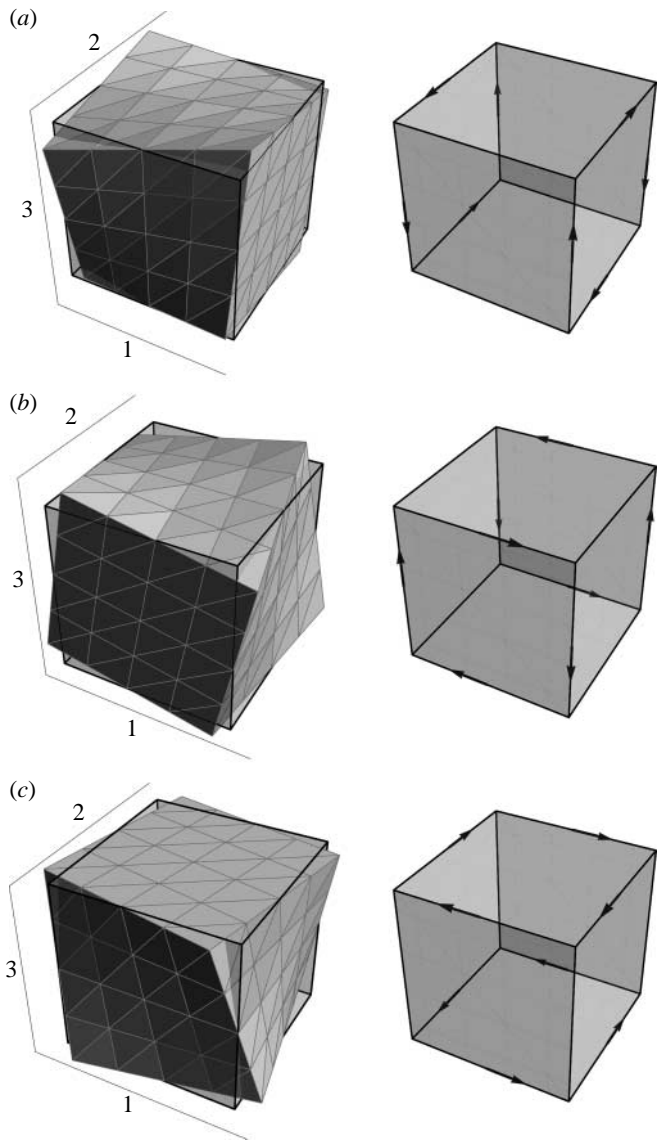

Figure 6. The second gradient deformation and hyperstress components: (*a*) $\hat{E}_{11}$ and $\hat{P}_{11}$; (*b*) $\hat{E}_{22}$ and $\hat{P}_{22}$; and (*c*) $\hat{E}_{33}$ and $\hat{P}_{33}$. These are coupled by $\mathbf{\Gamma}_2$ in equation (4.4)$_1$. Remark that $\hat{E}_{11}$, $\hat{E}_{22}$ and $\hat{E}_{33}$ are linearly dependent since the second-order tensor $\hat{E}_{ij}$ is deviatoric.

# Appendix A

A standard technique to visualize strain and stress fields is to display the deformations of an elementary reference volume (usually a cube) together with the associated contact forces on its boundary. Here, a similar approach is used to display the elementary states of strain-gradient and hyperstress, see equations (4.2)–(4.4).

The 'second gradient' deformations of an elementary cube centred in the origin are shown. We choose in the space of strain-gradients a suitable basis; for each element, say $C_{ijk}$, of this basis the corresponding displacement is chosen to be the second-order polynomial $u_i(X_1, X_2, X_3) = (C_{ijk} X_j X_k)/2$. With these choices the averages in the cube of all the (first-gradient) deformation fields are vanishing. Yet, the hyperstress fields are constant in the cube; thus, by equation $(2.26)_2$, their contribution to the contact forces $t$ is vanishing, and they are displayed only through the associated contact double-forces $\tau$ and edge-forces $f$ on the boundaries of the cube. Grey arrows are used to display the contact double-forces $\tau$, while black arrows are used for the edge-forces $f$ (figures 2–6).